# Molecular genetics of Aedes aegypti: historical advances and oversights

Moshe E Jasper[1]*, Thomas L Schmidt[2], Ary A Hoffmann[1]

[1] Bio21 Institute, School of BioSciences, University of Melbourne, VIC, Australia, 3052

[2] School of Life and Environmental Sciences, University of Sydney, NSW, Australia, 2050

*Correspondence: moshe.jasper@unimelb.edu.au

## Abstract

The yellow fever mosquito *Aedes aegypti* has a rich history of molecular genetic work but in recent years this has lagged behind rivals such as *Anopheles*, owing to the nature of control efforts focused on *Ae. aegypti* and the complex structure of its genome. However, *Ae. aegypti* has now entered the whole-genomic era of molecular genetics, with many whole genome sequences of individuals now publicly available for analysis. These resources enable an unprecedented view on several core genetic questions that have vexed research on the species for decades. They also create the possibility for new genomic analyses to be developed in the future. In this review, we consider the history of molecular genetic technology as it applies to *Ae. aegypti*, with a special emphasis on how emerging pictures of the genome both enabled and in some cases constrained the development of work on population genetics and insecticide resistance. We integrate historical and contemporary perspectives to diagnose gaps and biases within the *Ae. aegypti* literature that require further clarification and research. In particular, these highlight underrepresentation of the Indo-Pacific region in whole genome datasets relative to its dengue burden, and the overlooking of copy number variation in resistance studies owing to difficulties of sequence alignment in duplicated regions, particularly around the glutathione S-transferase epsilon class of genes.

**Box 1: The stages of molecular genetic research in *Ae. aegypti***

This review examines the historical progress of cytology, genomics and genetic marker work, looking at how their strengths and weaknesses directly affected investigations of population structure and resistance. We adopt a periodization loosely based on genomic marker type and the state of the genome, enabling us to study the species in the following overlapping eras:

1. **Morphological formal genetics (1958-1970)**

This period was marked by the identification of initial variants, the mapping of linkage groups and their identification within chromosomes, and the initial characterization of DDT and dieldrin resistance.

2. **Isozyme genetics (1969-1990)**

This period involved an initial exploration of molecular variation, leading to the first global characterization of *Ae. aegypti* diversity, which was heavily influenced by debates over the relative status of *aegypti* vs *formosus* and especially via their admixed range in Kenya. Resistance genetics expanded to identify three separate resistant loci, distinctions between larval and adult action as well as insensitivity versus metabolism

3. **Molecular genetics (1989-2007)**

This period involved the cloning and characterization of several resistance genes, culminating in the identification of knock-down resistance involving the *para* gene. It also involved the widespread use of RAPD markers and the steady progress of attempts to sequence the genome of the species.

4. **The genome-wide era (2007-2017)**

This era was dominated by the successes and complexities of working with the initial *Ae. aegypti* draft genome. Immediate benefits of its development included improved microsatellite panels, enabling much population genetic work. In addition, microarrays were developed targeting all gene families potentially involved in resistance, leading to a variety of transcriptomics tests. PCR-based genotyping of the voltage-sensitive sodium channel (*VSSC*) gene took off, with the discovery of a variety of mutants at this locus and their characterization with respect to pyrethroid resistance. The cytochrome P450 family rose to prominence in metabolic resistance research, with some attention to carboxylesterases and very limited attention to glutathione S-transferases (GSTs). As the decade wore on, double-digest restriction site associated DNA (ddRAD) sequencing and SNP chip technologies led to many studies using genome-wide SNPs, which improved resolution of population genetic studies but still suffered from poor mapping. Mapping improved around 2013-2014, and then in 2017-18 reached the standard required for the next era.

5. **The whole-genome era (2018-2026)**

The release of the *AaegL5* genome (Liverpool strain) provided for the first time physical mapping of most genetic sites to actual chromosomal positions was possible. Key developments at this time included the study of genome-wide CNVs in connection with resistance, as well as the establishment of the origin and global timings of *Ae. aegypti* subspecies emergence and introduction into Africa via widespread whole genome sequencing (WGS) datasets.

*Aedes aegypti* first sprang to global attention when it was identified as the principal vector of yellow fever (Finlay 1881, Reed et al. 1900), and later dengue fever (Ashburn and Caraig 1907, Cleland et al. 1916, Cleland and Bradley 1919), and when vector control efforts were shown to be effective at combating that disease (Gorgas 1901, Reed and Carroll 1901). At this time, research and policy focused on the reduction and later eradication of *Ae. aegypti* as a primary means of combatting its transmitted diseases. Control efforts generally followed the pioneering work in Havana, Cuba, which emphasized the reduction of breeding sources and their treatment with oil and larvicides, as well as the destruction of adult mosquitoes proximate to sites of infection through fumigation via the burning of pyrethrum powder or sulphur (Gorgas 1901, 1907). Improvements in pyrethrin usage (Tutin 1929, Barthel et al. 1944) and the development of DDT(Lindquist et al. 1945) led to large-scale efforts to eradicate the species from the Americas (Severo 1955).

The first examination of *Ae. aegypti* chromosomes occurred in 1918 (Carter 1918), but formal genetic study lagged for another four decades, other than the discovery that a strain could be selected for susceptibility to the parasite *P. lophurae* (Trager 1942). By the end of the 1950s, the anatomy, life history and physiology of *Ae. aegypti* had been extensively studied (Christophers 1960) but its genetic foundations were all but non-existent. The discovery of DDT resistance in *Ae. aegypti* (Brown 1956, Gilkes et al. 1956, Busvine and Coker 1958) and of its hereditary basis and capacity for selection (Shidrawi 1957, Coker 1958, Qutubuddin 1958) changed this situation entirely.

In two papers published 1957, P. F. Mattingly summarized the little that was known about *Ae. aegypti* genetics at the time, and, in the context of the emergence of insecticide resistance, issued a clarion call to the field: "*In the days when all that was attempted was control on a limited scale it was sufficient to attack the individual mosquito at the physiological level. Physiological homeostasis, however, has its counterpart in the genetically based homeostasis of populations, and when eradication or control over wide areas is desired it becomes necessary to envisage the species at a higher level of organization. This demands biology of an altogether more mature kind than has hitherto been applied to problems of mosquito control*" (Mattingly 1957a: 392, 1957b). In these seminal papers, he laid the foundations for two key pursuits that have shaped the field going forward: the genetic consequences of the global variability in *Ae. aegypti*, as seen especially in its subtypes (Mattingly 1957a), and the genetic bases of insecticide resistance (Mattingly 1957b). At the conclusion of his review, he called for enquiry into "*… the relationship between acquired insecticide resistance and the genetical structure of the particular population in which such resistance is found*" (Mattingly 1957b: 15). In the subsequent development of this field reviewed here, both of these aims are taken up and pursued with the available genetic resources of each era of molecular genetics (Box 1). But it is only comparatively recently that their integration has once again been attempted. A review of key achievements and ongoing challenges or oversights of each era follows.

### 1. Morphological formal genetics (1958-1970)

The first era of systematic research in *Ae. aegypti* was characterized by the search for morphological variants within existing laboratory strains, to identify traits for use in genetic mapping. Much foundational work occurred at the University of Notre Dame, where Prof. George Craig led a team that examined over 30 strains kept within their laboratories, along with adults which had been preserved periodically across the strains' histories (Craig et al. 1961). The work involved crossing of lines with detectable variants and calculation of gene frequencies within $F_1$ pair-matings to confirm the variants' Mendelian basis. In the five years following the launch of this research effort, more than 50 such variants were identified and described (Coker 1964).

A nascent understanding of the *Ae. aegypti* genome began in 1959, with the mapping of the mutations *yellow larvae (y)*, *black tarsi* (*blt*) and *black palp* (*blp*) to autosomal linkage groups assumed to correspond to chromosomes (Craig and Gillham 1959). As linkage group 1 was reserved for the sex-determining chromosome, *y* was assigned to linkage group 2, and *blt* and *blp* to linkage group 3. Following the discovery of the sex-chromosome-linked mutation *red eye* (*re*) during George McClelland's PhD research, triple-crosses led him in 1962 to the first genetic map of the linkage groups as well as to the development of the *triple-marked strain*, with the markers *re* (group 1), *spot* (group 2) and *blt* (group 3) widely used in later mapping experiments (McClelland 1962). A more comprehensive map published in 1967 contained a total distance of 110 crossover units but was as yet unconnected to physical chromosomes (Craig and Hickey 1967).

Feulgen staining of *Ae. aegypti* chromosomes by Rai in 1963 led to the physical identification of the chromosomes by size and labelling as (in ascending order): I (the sex chromosome), II, and III (Rai 1963). However, these physical chromosome designations were not connected to the linkage groups until radiation experiments successfully induced two reciprocal translocations, one between linkage groups 1 and 2, and the other between groups 1 and 3 (Rai et al. 1970). These cytologically visible translocations led in 1970 to the successful correlation of chromosomes and linkage groups (McDonald and Rai 1970). This led to a redesignation of chromosomes III and II (defined by length order) as chromosomes 2 and 3, now corresponding to the order of identified linkage groups rather than length order. This crowning achievement led by the close of this era to a chromosomal linkage map which containing 28 loci including three associated with either DDT or dieldrin resistance (see section 1.2), the identification of around 87 morphological mutants (Craig and Hickey 1967), and the establishment of marker strains to support ongoing research.

### 1.1 Population genetics – Morphological formal era

During this era, population genetics of wild populations was constrained by the necessity of strain crossing to establish the Mendelian basis of morphological traits. Early work involved surveying morphological variation in laboratory strains of differing geographical origin (Craig et al. 1961). Wide variation was found in the distribution of the yellow larvae gene, *y*, across 25 strains sourced from thirteen countries, as well as in that of other traits. However, these were still for the most part laboratory strains which had been maintained in captivity for years. Around 1962, McClelland examined pigmentation from 35 strains newly established from field locations and examined the frequency of colour variants. He identified two additive major alleles controlling the scaling of the abdominal dorsum, underlaid by further variance that was likely polygenic (McClelland 1962). His work also led him to challenge the usefulness of the variant *queenslandensis*, while upholding the probability of a *formosus* subspecies, provided it was interpreted with respect to both behaviour and colour (McClelland 1962) – a conclusion which Mattingly also reached (Mattingly 1967). This work was subsequently extended to samples from 33 countries (McClelland 1974). Basic tests of heterozygosity were conducted from five African localities via $F_1$ pair mating and identification of mutants in the $F_2$ generation. Three “feral” West African populations (Ghana, Nigeria, Rep. de Congo) were found to exhibit lower average mutation count per mosquito (0.52–0.84) than two “domestic” populations from Kenya (2.72–2.96) (Craig and Hickey 1966). A previous survey of laboratory strains had found 1.32 mutations per mosquito (Vandehey 1964). Overall, population genetics in this era made little progress, with key achievements being the recognition of genetic diversity and challenges to the distinct status of *Ae. aegypti* v. *queenslandensis*. An additional effect of these initial investigations was to centre subsequent research attention on domestic *Ae. aegypti* populations in Kenya.

### 1.2 Resistance genetics – Morphological formal era

By 1957, DDT resistance had been detected in Trinidad, Dominican Republic, Colombia & Venezuela (Brown 1958), and resistance to both DDT and dieldrin was also detected in a 1952 strain from Selangor, Malaysia (Shidrawi 1957). Larval DDT resistance in both regions was associated with detoxification of DDT to DDE via dehydrochlorination in a reaction involving glutathione (Brown 1967).

The first explicitly genetic experiments on resistance started with the M.Sc. work of William Coker, who undertook crossing experiments and larval assays involving Caribbean and Malaysian strains, concluding that resistance in each was likely driven by a single factor (Coker 1958). Selection-relaxation experiments on another cross-resistant wild-caught strain from Penang, Malaysia showed swift development of likely multi-factorial resistance to DDT, with initial costs that reduced as selection proceeded (Abedi and Brown 1960). Resistance characteristics in this Malaysian population were different to Caribbean strains (Busvine and Coker 1958). Following the establishment of linkage groups, DDT ($R^{DDT1}$) and dieldrin ($R^{Dl}$) resistance from the Isla Verde strain, Puerto Rico, were both due to a single factor mapping to linkage group 2 (Khan and Brown 1961). Initial crossing experiments with an additional group 2 marker (*spot*) and three more resistant strains showed that multiple resistance profiles (malathion, DDT-only, DDT and dieldrin cross-resistance) from multiple origins (Malaysia and the Caribbean) traced to a similar location on linkage group 2, suggesting multiple allelism within the same cistron region (Brown and Abedi 1962). However, subsequent experiments revealed differential linkage patterns among sub-strains, with the Penang (Malaysia) strain and one Trinidad sub-strain placed at 16 crossover units from the *s* marker, while the other two Trinidad strains were placed at 26 units (Coker 1964). As the three Trinidad sub-strains were derived from the same initial selection, Coker suggested that multiple sites of resistance were present in the original stock, and that there was an additional influence of genetic background (Coker 1964). A dieldrin resistance site ($R^{Dl}$) was found to be separated from DDT resistance ($R^{DDT1}$) by crossover distances of 1–4% in additional strains from the Caribbean (Puerto Rico, Jamaica, Curacao) (Klassen and Brown 1964), and both were placed on the first linkage map (Craig and Hickey 1967). DDT resistance in Thai and West African lines was later found to map similarly to Trinidad and Penang, and dieldrin resistance from Pakistan and the Upper Volta (modern Burkina Faso) corresponded to the previously identified location (Inwang et al. 1967). An additional DDT resistance-enhancing site associated with malathion selection, $R^{DDT2}$, was located on linkage group 3 (Pillai and Brown 1965), and in further experiments with a Trinidadian sub-strain, the previously discovered DDT marker in linkage group 2 ($R^{DDT1}$) was associated with larval resistance, while $R^{DDT2}$, the major gene driving adult resistance, was found on linkage group 3, 18.2 units from *blt* (Wood 1967). These two DDT loci were corroborated in experiments on a Thai strain (Margham 1970).

Overall, the mapping of $R^{Dl}$ and $R^{DDT1}$ to chromosome 2 and $R^{DDT2}$ to chromosome 3 were key achievements of this era. Interestingly, at this time DDT resistance on chromosome 2 was seen as the major effect, and that on chromosome 3 as secondary. While metabolic and insensitivity resistance mechanisms were both recognized, as well as distinctions between adult and larval action, detoxification was assumed to make the major contribution to resistance. Importantly, the dominant mechanism of initially detected DDT resistance in larval experiments involving both Malaysian and Caribbean resistant strains was associated with the *R^DDT1* locus on chromosome 2(Khan and Brown 1961) and shown to require the presence of glutathione (Pillai and Brown 1965, Brown 1967). This suggests the distinct possibility that sweeps involving glutathione S-transferase epsilon class (*GSTe*) genes from chromosome 2 were involved in early resistance evolution. At the same time, *R^DDT2*, later recognized as the voltage-sensitive sodium channel (*VSSC*) gene, was additionally implicated as acting in Thai and Caribbean resistant strains, especially in adult resistance.

### 1.3 Challenges – Morphological formal era

A core challenge experienced in this era which continues to impact *Ae. aegypti* genomics was the difficulty of working with physical chromosomes. In her initial work on *Ae. aegypti* chromosomal structure, Lucy Carter erroneously described the diploid number of chromosomes as four, but noted the a need for "precaution" in sectioning when working with the species' long chromosomes, and stated, "in the large cells, where the chromosomes are very long ... they are often difficult to count" (Carter 1918). Her statement that "most probably because of this great length the usual V-shaped chromosome of anaphase is often replaced in *Stegomyia* by a W-shape" suggests the close association of chromosomal pairs contributed to this undercount. While her later work on (polytene) salivary gland chromosomes corrected Carter's estimate, Eileen Sutton noted that standard squash preparations, rather than spreading *Ae. aegypti* chromosomes out, "merely squashes the whole nucleus" (Sutton 1942). The banded structure used to map other polytene chromosomes could not be consistently followed as chromosomal arms were riddles with "weak spots" where chromosomal arms would fragment and parts of the same or different chromosomes would adhere to the break, obscuring its continuity. In sharp contrast to contemporary work on *Anopheles* mosquitoes (Frizzi 1953, Rioux et al. 1958), polytene chromosomes proved difficult to work with and were rarely able to be uncoiled enough to allow band mapping (Aldighieri 1961). This situation was to hold back *Ae. aegypti* genetics and then genomics all the way until around the 2010s, where advances in alternative physical mapping techniques (Sharakhova et al. 2011, Timoshevskiy et al. 2012, 2013) and sequencing technologies (Dudchenko et al. 2017) finally allowed the detailed mapping of physical and genetic material.

## 2 Isozyme genetics (1969-1990)

The first application of protein electrophoresis to mosquitoes occurred in 1968, when twelve species of mosquitoes including *Ae. aegypti* were examined for patterning of six enzymes: acetate and butyrate esterases, alkaline and acid phosphatases, and malic and lactic dehydrogenases (Trebatoski and Haynes 1969). In subsequent work, strain variation within *Ae. aegypti* was examined using alpha-naphthyl acetate, and six bands were identified, corresponding to esterases, of which two were allelic. The fastest-migrating, *Esterase 6,* showed differential migration in three lines, and reciprocal crosses showed co-dominance (Trebatoski and Craig 1969). It was mapped to linkage group 2, 17.4 units from *yellow*. By 1976, fourteen alleles had been identified at the esterase-6 locus (Saul et al. 1976). By 1979, a linkage map incorporating both 46 phenotypic markers and 14 isozyme markers was produced (Munstermann and Craig 1979).

Several cytological advancements further improved broad-scale understanding of the *Ae. aegypti* genome. First, studies of chiasmata frequencies led to an estimation of 228 map units within the species (Munstermann and Craig 1979). Second, further experiments inducing translocations via X-irradiation led to the physical mapping of the *black tarsi* (*blt)* locus to a terminal segment of the left (small) arm of chromosome 3 distal to the arm exhibiting a secondary constriction, establishing a definite orientation to this chromosome's map (Bhalla 1973). Recently developed Giemsa C-banding enabled the X and Y chromosomes to be distinguished, as the Y chromosome's centromere did not stain (Newton et al. 1974). C-banding was also successfully used to differentiate *Ae. aegypti* and *Ae. mascarensis* and their hybrids (Motara and Rai 1977), and to map the sex-linked *distorter* gene to between the centromere and intercalary band (Newton et al. 1978). These studies established the difference between constitutive and facultative heterochromatin in the species, and suggested a role of the latter as well as some chromosomal rearrangements in *Aedine* speciation (Rai 1980).

### 2.1 Population genetics – Isozyme era

The arrival of codominant isozyme markers occurred during a critical phase in the investigation of the genetic structure of *Ae. aegypti*. At this time, research attention had turned towards understanding the basis of the variation in forms within *Ae. aegypti*, amidst an ongoing debate about the taxonomic status of *Ae. aegypti* s. *formosus (Aaf)* and *Ae. aegypti* s. *aegypti (Aaa)*. This has been prefigured by McClelland's dismissal of type *queenslandensis* in 1962 – although he was more accepting of type *formosus,* writing, "*The occurrence of some forms as dark as ssp. formosus in the Caribbean, Indonesia, etc. and some paler forms in the heart of Africa need not lessen its value as a subspecies, providing the non-African forms show the peri-domestic habit and the paler African forms retain feral behaviour-patterns. On the other hand, as has been suggested above, the definition of Mattingly (1957) was a little too stringent"* (McClelland 1962: 216–217). By 1976, the alternatives presented were subspecies (Mattingly 1967) and ecotype (Scott and Mcclelland 1975), alternatives which prefigure similar discussions today (Powell et al. 2025, Wilkerson and Harbach 2025 July 1). Much of this work centred on Rabai district, Kenya, where light-coloured domestic and dark-coloured sylvan forms were sympatric (Tabachnick et al. 1979). Experiments at this time showed dark *Aaf* individuals were much less anthropophilic than light *Aaa* (including *queenslandensis*) individuals (Gouck 1972), as well as more susceptible to desiccation (Machado-Allison and Craig 1972). However, both interbred readily and without cost in the laboratory (Moore 1979). Mark-release-recapture (MRR) experiments in Rabai, Kenya found behavioural differences between domestic, peridomestic and urban mosquitoes, with more domesticated mosquitoes being recaptured more frequently within houses (Trpis and Hausermann 1975). A likely genetic basis to this was found via further MRR experiments involving crosses between the differing strains (Trpis and Hausermann 1978).

Population genetic work on this question was advanced considerably when a six isozyme system was applied to *Aaa* and *Aaf* individuals in Kenya, showing differences in gene frequencies between populations of each type consistent with reduced gene flow (Scott and Mcclelland 1975). A separate study using the Est-6 isozyme showed both great diversity across globally sourced strains, as well as differences in gene frequency between both indoor and outdoor populations within a village and from one local village to the next in Rabai, Kenya (Saul et al. 1976). Indeed, if was shown that *Aaf* populations from across East Africa more closely matched isozyme frequencies of *Aaf* in Rabai, Kenya than *Aaf* and *Aaa* matched within Rabai (Petersen 1977). In his PhD thesis, Peterson first proposed the origination of *Aaa* from *Aaf*, its separation and adaptation to humans in the arid climate of West Africa, and the subsequent arrival of *Aaf* into coastal Kenya, a region already inhabited by *Aaa* (Petersen 1977). This position substantially connects with recent accounts (Rose et al. 2023).

Subsequent to the use of isozymes in Kenya, in 1979, the first truly global population-genetic study in *Ae. aegypti* was published involving 34 populations and 19 isozyme loci (Tabachnick and Powell 1979). This study re-used the previous Kenyan populations and added populations from Tanzania, the Upper Volta (Burkina Faso) and Nigeria in Africa (21 populations), Jamaica, Puerto Rico, Mexico, Venezuela and the U.S.A in the Americas (9 populations), and Indonesia, India and Taiwan in the Indo-Pacific region (4 populations). No *Aaa* were collected from Africa outside of Rabai district, Kenya. Indo-Pacific *Aaa* populations were found to be significantly less heterozygous that African sylvan populations (*Aaf*) or Americas domestic (*Aaa*) populations. Dendrograms based on Nei's distance (Nei 1972) clustered Americas and East Africa *Aaa* as distinct from *Aaf*, but grouped Indo-Pacific *Aaa* close to East African *formosus*. This paper proposed multiple introductions of *Ae. aegypti* into the New World as an explanation of the heterozygosity found there, and a separate introduction of *Ae. aegypti* into the Indo-Pacific. It should be noted, however, that this study is based on a limited number of isozymes, and that the authors reasoning presumed that East African *Aaa* populations were present prior to the global invasion. No *Aaa* from West Africa were included, and very few populations from the Indo-Pacific.

Using the same dataset, Powell et al. (1980) applied a multivariate discriminate analysis and confirmed clusters corresponding to the Americas, African *Aaf* and the Indo-Pacific, with African East-coast *Aaa* clustering with the Americas. They noted that the Indo-Pacific appeared intermediate between Africa and the Americas because allele frequencies in Asian populations resembled those of *Aaf* at the *Idh-2* locus (Powell et al. 1980: 208, Table 2). This illustrates the inconsistencies across loci that can arise from using isozyme methods. The same isozyme methods were applied to explore population structure in New Orleans and Florida (Tabachnick 1982), the Caribbean (Wallis et al. 1984) and Houston, Texas (Harrington et al. 1984). By 1983, global population genetics had extended to 63 localities, albeit at only 11 enzyme loci (Wallis et al. 1983). The work extended the initial 1979 dataset to include populations from Senegal, Ghana and Gambia in Africa, from Guyana, Suriname, Guatemala, Trinidad and additional U.S.A. populations in the Americas, and from Fiji in the Indo-Pacific (Wallis et al. 1983, Table 1). UPGMA-based distance trees suggested south-eastern U.S.A. and the Indo-Pacific to be most closely related to *Aaf* forms. However, this was once again driven by *Idh-2* frequencies. Once East African *Aaa* subspecies were removed, canonical analysis in fact shows Indo-Pacific populations more closely associated with the Americas than with Africa (Wallis et al. 1983, Figures 4-5). Additional isozyme studies occurred in a variety of locations over the succeeding decades (Failloux et al. 1995, Lerdthusnee and Chareonviriyaphap 1999, Tien et al. 1999, de Sousa et al. 2000, Paupy et al. 2000, 2003, 2004, A.-B. Failloux et al. 2002, Huber, Luu Le Loan, et al. 2002, Mousson et al. 2002, Yébakima et al. 2004, Sukonthabhirom et al. 2005, 2009, Herrera et al. 2006, Huber et al. 2008, Santos 2011). The last major global synthesis of population genetics involving isozymes identified three global clusters: a sylvan form from West Africa and islands in the Indian ocean, a domestic form from South-East Asia and South America, and a domestic form from the South Pacific islands (Anna-Bella Failloux et al. 2002). Its findings stand in partial contradiction with the previous global study, grouping the Americas with South-East Asia. However, the dataset included more Indo-Pacific samples and did not include North American or East African samples.

### 2.2 Resistance genetics – Isozyme era

Resistance studies continued to clarify the distinctions between the three major loci identified in the previous decade. By 1970, studies were divided between those finding a major DDT resistance gene on chromosome 3(Wood 1967, Margham 1970) via adult knockdown assays, and those instead finding a major DDT resistance gene or genes on chromosome 2 via larval assays (Khan and Brown 1961, Brown and Abedi 1962, Coker 1966). Following further experiments on Thailand strains, Margham and Wood (1975) suggested that resistance was being jointly driven by two genes of major effect. These were $R^{DDT2}$ on chromosome 3, and $\mathrm{R}^{\mathrm{DDT1}}$ on chromosome 2, with the former having the major effect in their resistant strain, and the latter becoming more important when they outcrossed with susceptible mosquitoes and re-selected for resistance. The gene $R^{DDT1}$ was suggested to be the primary locus associated with larval DDT resistance. Extending the experiments to strains from Trinidad (Caribbean) and Bangkok (Thailand) found for both strains the major locus in adult DDT resistance was $R^{DDT2}$ (chromosome 3), with crosses between strains exhibiting allelism and mapping to the same chromosomal region, 19 units from *blt* (Hitchen and Wood 1975a). Cross mapping of larval assays for $R^{DDT1}$ (chromosome 2) found that Bangkok and Trinidad resistance here was allelic and well separated from dieldrin resistance from the Isla Verde strain at $R^{dl}$ (Hitchen and Wood 1975b). A separate Bangkok strain showed influence of linkage group 2 on adult DDT resistance (Margham and Wood 1976).

At this time, cross resistance to both pyrethroids and DDT had been found in several mosquito species, including *Ae. aegypti* (Rongsriyam and Busvine 1975). Cross-resistance to DDT and pyrethroids was confirmed in strains from Bangkok and Jakarta, South-east Asia, as well as from the Caribbean and central Africa (Chadwick et al. 1977). Cross-resistance investigations in Central and South American strains identified a dehydrochlorination mechanism in highly resistant Guyana and Salvador strains, with an additional role for microsomal oxidase systems in Guyana (Prasittisuk and Busvine 1977). Other contemporary studies identified three contributing factors to larval resistance: (1) increased DDT metabolism via dehydrochlorination; (b) increased tolerance to unmetabolized internal DDT; and (c) reduced uptake of DDT (McDonald and Wood 1979). Further work on global strains found strains of Asian and West African origin showed DDT resistance as larvae but not adults, while strains examined from the Americas showed resistance in both stages (Rathor and Wood 1981). Resistant strains from the Americas showed increased DDT dehydrochlorination as larvae.

The next step came with the establishment of a homozygous, highly permethrin-resistant strain via selection of a cross-resistant mosquito line (Chadwick et al. 1977) initially collected in Bangkok, Thailand (Malcolm and Roger J. Wood 1982). Using this strain, the major gene of pyrethroid knockdown resistance, $R^{py}$ was mapped to chromosome 3, 13 map units from *black tarsus* (*blt*) (Malcolm and R. J. Wood 1982), and shown to be allelic to chromosome 3 $R^{DDT2}$ (Malcolm 1983a, 1983b). Accordingly, this study renamed the chromosome 2 resistance gene $R^{DDT}$ and the chromosome 3 gene $R^{DDT\text{-}py}$. This represents the first genetic mapping of cross-resistance: "*the relationship between DDT resistance and permethrin resistance in adult Ae. aegypti is therefore apparently simple. Two major unliked DDT resistance genes can be present; one, $R^{DDT}$, on chromosome II controls the resistance mechanism dehydrochlorination and confers a level of resistance to DDT 3-4X higher than the other, but produces no cross-resistance to permethrin. $R^{DDT2}$, henceforth termed $R^{DDT\text{-}py}$ isolated in a susceptible background, confers resistance to DDT of about 10-14X and cross-resistance to permethrin of 18-21X*" (Malcolm 1983b: 228). Further metabolism research in these strains suggested that pyrethroid resistance was of the *knockdown* (*kdr*) type, while that of DDT was a mixture of *kdr* and DDT-dehydrochlorinase (Brealey et al. 1984).

### 2.3 Challenges – Isozyme era

While isozymes enabled the molecular characterization of heterozygosity and reduced gene flow in field populations for the first time, they suffered from several limitations, including an insufficient number of markers, leading to very low resolution in phylogenetic separation compared to later methods (Bossart and Prowell 1998). The low number of markers in conjunction with the fact that these loci affect the behaviour of proteins make them additionally susceptible to selection driving allele frequencies away from neutrality (Allendorf and Seeb 2000). As isozyme studies often only tend to use markers that are at high frequencies, they have been challenged as at particular risk of identifying alleles that have risen to prominence due to selection (Mani et al. 1986). Indeed, isozyme studies are often strongly influenced by the exceptional variability of one or two alleles which have undergone selection (Allendorf and Seeb 2000). These factors have had a definite impact on the isozyme-based investigation of population structure in *Ae. aegypti*. Both major early global isozyme studies (Tabachnick and Powell 1979, Wallis et al. 1983) identified Indo-Pacific populations as apparently intermediate between African and American populations. However, this was driven almost entirely by the high contribution of a single isozyme locus, Idh-2, whose gene frequencies, in contrast to other markers, resembled those of *Aaf*.

## 3 Molecular genetics: 1989-2007

The molecular genetics era involves the move to the fore of DNA-based technologies such as fragment amplification and cloning. This phase of *Ae. aegypti* genetics began with the cloning, sequencing and identification of the *Maltase-like I (MalI)* gene from cDNA libraries of the salivary glands of adults (James et al. 1989). Following sequencing, expressed probes were used to identify the sequence within a library of genomic clones, enabling the identification of the transcription start point as well as a 3 kb intron near the 3' end of the gene. At approximately the same time, *Ae. aegypti* rRNA was identified and characterised through restriction and transcription mapping as 9.0 kb in length with high homogeneity, comprising 0.6% of the genome (Gale and Crampton 1989). Later, families of transposable elements were characterized (Tu 1997), and eventually, in 2002, the complete mitochondrial sequence was published (Morlais and Severson 2002).

The success of this emerging molecular technology led to the goal of characterizing the *Ae. aegypti* genome, which would define the next decade or more of research. But this goal needed to meet a variety of challenges. Initial analysis via reassociation kinetics revealed that the genome was large, complex, and dominated by short interspersed repeats (Warren and Crampton 1991). This was immediately recognized as a challenge to exploratory techniques such as clone hybridization. In addition, the low variability of mtDNA and presence of rDNA sites on autosomes served to reduce the usefulness of these DNA markers (Besansky and Collins 1992). While polytene chromosome microdissection and amplification were transforming *An. gambiae* genomic mapping (Zheng et al. 1991), *Ae. aegypti* polytene chromosomes continued to be unsatisfactory, "probably owing to the fact that the abundant highly repetitive dispersed DNA sequences promote extensive ectopic pairing" (Besansky and Collins 1992). Instead, a new linkage map was constructed using restriction fragment length polymorphisms (RFLPs) (Severson et al. 1993). This map contained 50 DNA markers covering 134 map units, but was only partially integrated into existing genetic maps, with only around 10 previous markers incorporated.

In an attempt to overcome the mapping issues owing to the failure of polytene chromosomes, fluorescence in situ hybridization (FISH) was applied to the metaphase chromosomes of an *Ae. aegypti* cell line, mapping 37 probes (Brown et al. 1995). Of these, twenty-three were subsequently developed into expressed sequence tags (Severson and Zhang 1996). Further FISH developments enabled chromosomal landmarks to be identified which specify a chromosome and its orientation, facilitating physical mapping (Brown and Knudson 1997). Initial physical mapping was integrated with genetic maps through connecting RFLP markers to cosmids and assigning these to chromosomal locations through FISH (Brown et al. 2001). This mapping showed that large regions of chromosomes 1 and 3 were not covered by existing genetic maps. It was estimated that only 45% of the genome was transcriptionally active.  A further attempt to spread polytene chromosome was made in 2003 (Campos et al. 2003).

A separate linkage map was constructed with 96 RAPD markers, but only one genetic marker per chromosome was incorporated, and no RFLP markers (Antolin et al. 1996). The reliance on dominance markers inherent to this method reduced its usefulness, and many of the markers were not found in other *Ae. aegypti* populations (Bosio et al. 2000). A new map was developed via single-strand conformation polymorphism (SSCP) analysis of genes with known function, leading to a linkage map involving 53 genes (Fulton et al. 2001). However, these maps were predominantly rendered redundant by an updated RFLP map, also incorporating some SSCP markers, which consisted of 145 sequence-tagged site (STS) loci distributed across 205 cM (Severson et al. 2002). This map was explicitly published to support genome sequencing efforts in the wake of the publication of the *D. melanogaster* genome (Adams et al. 2000), and included sequence data for all mapped clones. The STS tags associated with this map were then used in the construction of a bacterial artificial chromosome (BAC) library to support physical mapping efforts (Jiménez et al. 2004).  Overall, this era, commencing with the cloning and isolation of genes, was defined by a series of mapping and sequencing efforts culminating in the publication of the first draft genome for *Ae. aegypti* in 2007 (Nene et al. 2007).

### 3.1 Population genetics – Molecular era

Population genetics studies between 1990 and 2007 primarily used DNA-based markers to examine population structure across local or regional scales. The first DNA-based population genetics experiment used random-amplified polymorphic DNA (RAPD) markers (Ballinger-Crabtree et al. 1992), and examined 11 populations of *Ae. aegypti*. Most populations were from established colonies rather than field sampling. Dendrograms based on genetic distance, similarity index and percent match showed clear separation between *Ae. a. aegypti* and *Ae. a. formosus*. Nearest neighbour discriminant analysis was used to classify individuals by population. The increased variation captured by RAPD markers allowed for additional applications, such as estimating the number of full-sibling families in a mixed group (Apostol et al. 1993), and analysing breeding structure via $F_{ST}$ and θ across Puerto Rico (Apostol et al. 1996). The development of RFLP-based genetic maps indicated that they could help determine genetic relatedness among field populations (Severson et al. 1994 Apr 1). Both RFLP and amplified fragment length polymorphism (AFLP) were used to examine population genetic structure in Trinidad (Yan et al. 1999). It was shown that AFLP and RAPD markers, involving simple Mendelian dominance, were likely to lead to bias in estimates of genetic variability within and between populations. Subsequent RAPD studies examined populations in Mexico (Gorrochotegui-Escalante et al. 2000), Argentina (de Sousa et al. 2001), Brazil (Ayres et al. 2003, Santos et al. 2003, Paduan et al. 2006, de Patarro et al. 2013 Sept 19), Colombia (Ocampo and Wesson 2004), Pakistan (Ashraf et al. 2016) and India (Gupta et al. 2021).

The first mitochondrial haplotype, NADH dehydrogenase subunit 4 (*ND4*), was used in an isolation-by-distance study of north-eastern Mexico, in an attempt to corroborate RAPD methods (Gorrochotegui-Escalante et al. 2000). In this study, genetic and geographical distances were correlated using RAPDs, but not using SSCP-genotyped mtDNA. A follow-up study examined *ND4* haplotypes across Mexico, using $F_{ST}$ to infer genetic structure between populations in Northeastern Mexico, the Yucatan and the Pacific (Gorrochotegui-Escalante et al. 2002). Another study in Venezuela using 1,144 *Ae. aegypti* from 24 sites in eight regions used AMOVA to infer significant differentiation at all study scales (Herrera et al. 2006). Work on this mitochondrial locus culminated with a global analysis of 36 locations across the Americas (Brazil, Peru, Venezuela, Guatemala, USA), Africa (Guinea, Senegal, Uganda) and the Asia-Pacific (Singapore, Cambodia, Tahiti) (Bracco et al. 2007). The study identified 20 haplotypes, of which 18 were confined to individual regions. Two deep clades were present in the Americas, which the authors interpreted to indicate either recent global contact or a broader failure of eradication efforts. Additional studies using this locus were carried out in the coming years (Paupy et al. 2008, 2012, Urdaneta-Marquez et al. 2008, Gonçalves da Silva et al. 2012, Moore et al. 2013, Muñoz et al. 2013, Yáñez et al. 2013, Aguirre-Obando et al. 2015, Jaimes-Dueñez et al. 2015, Ali et al. 2016, Calvez et al. 2016, de Sousa et al. 2017, Salgueiro et al. 2019, Seixas et al. 2019, Campos et al. 2020, Fernando et al. 2020, Khater et al. 2021).

In 2009 the *Ae. aegypti* genome was shown to contain many nuclear copies of mitochondrial genes, including the *ND4* gene (Hlaing et al. 2009). The two highly divergent lineages seen in this gene showed composite presence in many samples, indicating nuclear amplification, leading the authors to urge avoidance of the *ND4* marker, and take caution with mitochondrial markers in general. An alternative mitochondrial marker, the *CO1* gene, was used to study the origin of an invasive *Ae. aegypti* population in Tennant Creek, Australia (Cooper et al. 2005). Haplotypes found in Tennant Creek matched those from Cairns but differed from those in Townsville (both sites in Queensland, Australia). This haplotype, H1, was also found in Southeast Asia, and was closely related to another haplotype found in Indonesia, Myanmar, Vanuatu and PNG (H4).

In the early 2000s, several microsatellite markers were developed. Without the genome assembly as a guide, the identified markers lacked diversity and were limited in power. For example a study using four AFLP markers and three microsatellites examined the genetic diversity of two north-west Mexican towns (Ravel et al. 2001). These microsatellites were of poor quality and unable to distinguish mosquitoes from either town, with only five segregating alleles across two loci. AFLP markers were applied to populations to estimate diversity statistics; however, many individuals showed null amplification for the markers. Another study using six microsatellite loci (5-14 alleles per locus) was able to differentiate populations from the city centre of Ho Chi Minh City from those in the outskirts via $F_{ST}$ (Huber, L. Le Loan, et al. 2002), showing improved performance relative to a prior isozyme study (Tien et al. 1999). Isozymes, AFLP and microsatellites were compared in a study in Cambodia, with AFLP showing similar population structure but much higher $F_{ST}$ values (Paupy et al. 2004).

### 3.2 Resistance genetics – Molecular era

#### *3.2.1 Dieldrin resistance (Rdl)*

The molecular basis of dieldrin resistance was discovered via cloning and sequencing of the cDNA homologous to the *Drosophila* resistance gene *Rdl* (Thompson et al. 1993, Feyereisen 1995). The same amino acid substitution in the channel lining of γ-aminobutyric acid gated chloride ion channel (GABA) that had been seen in resistant

*Drosophila* was also found in resistant *Ae. aegypti*. This represented the first resistance-associated substitution found, and was followed up with the development of a diagnostic PCR for the alanine to serine substitution A296S (Ffrench-Constant et al. 1994) and functional expression experiments which showed a 100-fold increase in insensitivity relative to wild type (Shotkoski et al. 1994).

Subsequently, a fragment of the acetylcholinesterase (*Ace)* gene was cloned via a degenerate primer strategy based on conserved insect sequences, identifying a 308-bp sequence which was used to identify the cDNA and derive its coding sequence (Anthony et al. 1995). While no resistance had been found in AChE at that time, site-directed mutagenesis indeed conferred resistance to organophosphates and carbamates (Vaughan et al. 1997). An RFLP-based examination of Trinidad strains collected in 1995 from populations subject to more than two decades of insecticide resistance showed a putative selective sweep, with reduced variation and increased $F_{ST}$ in the vicinity of the *Est*-4 gene (Yan et al. 1998).

### *3.2.2 Para (kdr)*

Little substantial work occurred on the $R^{DDT\text{-}py}$ cross-resistance gene from 1990 onwards, other than its identification with *para* from *D. melanogaster* and incorporation into RFLP genetic maps (Severson et al. 1997). Following identification of *kdr* mutants in this gene in other insects in the late 90s, an RT-PCR approach was used to obtain and sequence part of the S6 hydrophobic segment of domain II of the *VSSC* gene across 13 field strains, 10 of which showed larval permethrin resistance (Brengues et al. 2003). 407 bp were amplified, and six pyrethroid-resistant strains showed nucleotide substitutions resulting in the amino acid changes L982W (Long Hoa, Vietnam), G923V and I1011M (Brazil, French Guiana, Martinique) and V1016G (Thailand, Indonesia). They associate the first mutational types with larval insensitivity to pyrethroids.

### *3.2.3 Glutathione S-transferases (GSTs)*

Two *GST*s were characterized in permethrin and DDT resistant and susceptible strains of *Ae. aegypti*. The enzyme GST-*2* [now GST*x2*] (28 kDa, isoelectric point < 5.0) showed eight-fold higher activity throughout development in resistant strains compared to a susceptible strain (Grant and Matsumura 1989). GST enzyme activity was measured with CDNB (1-chloro-2,4-dinitrobenzene) and GSH (glutathione). A lack of correlation between DDT resistance and enzyme activity was noted. However, the measure of enzyme activity used was a general GST activity, not specific to DDT. Later work linked this higher enzyme activity to over-expression controlled by a *trans*-acting regulatory locus repressing *GST-2* [*GSTx2*] transcription or stability along with additional loci (Grant and Hammock 1992). It is important to note that at the time of these studies, only three GST enzymes were known, and the Epsilon family had not yet been identified.

Following the development of RFLP genetic maps, crossing experiments determined that the GABA receptor gene and *dieldrin* resistance map to the same genomic region on chromosome 2, as do the *para* gene and $R^{DDT\text{-}py}$ on chromosome 3 (Severson et al. 1997). As *para* had been identified for target site insensitivity in other insects, the researchers suggested the previously studied $R^{DDT}$ gene on chromosome 2 must be associated with metabolic activity, either GST or cytochrome P450.

### 3.3 Challenges – Molecular era

An important issue that emerged at this time was that new methods were often incompatible with previous existing work. While the identification of the genes associated with dieldrin and knockdown resistance were major achievements, their strengths are found in the new ability to apply fundamental experimental information (the mRNA sequence) from related organisms. Very little fundamental resistance characterization or mapping occurred in *Ae. aegypti* itself during this time period. Indeed, the move to DNA-based mapping often meant that entirely new chromosomal linkage maps were constructed with next to no points of reference with each other or with earlier genetic maps, with little in common other than the basic markers of the *triple-marked strain* (Antolin et al. 1996). In practice, this meant that any mutations not immediately to hand or of high importance were left out of DNA-based maps. So, for example, the initial RFLP map (Severson et al. 1993) contained only recently cloned genes and the *triple-marker* strain sites. While a subsequent study explicitly sought to insert resistance genes into this map, its primary interest was the recently cloned *Ace-1*, *Rdl* and *para* genes, all potential sites of target site insensitivity resistance (Severson et al. 1997). The chromosome 2 $R^{DDT}$ locus was included on the phenotypic map but not on the key composite map the paper is producing and is quietly discounted in the discussion as probably associated with increased metabolic activity.

## 4 The genome-wide era: 2007-2017

In 2007, a significant milestone in *Ae. aegypti* genetics was reached with the publication of its draft genome sequence (Nene et al. 2007). This followed the publication of the draft *Drosophila melanogaster* genome in 2000 (Adams et al. 2000) and that of *Anopheles gambiae* in 2002 (Holt et al. 2002). The genome assembly was 1,376 Mbp, substantially larger than previous estimates, and was spread across more than 1,257 scaffolds with an N50 size of 1.5 Mbp. Limitations in genetic and physical maps meant that only 238 scaffolds (31% of the genome) could be assigned to physical chromosomes, and these were assigned to chromosomal arms 'without order or orientation' (Nene et al. 2007: 1719). About half the genome was found to be transposable elements, leading to a four to six-fold increase in gene length relative to *An. gambiae* and *D. melanogaster* due to increased intron lengths. While transposable elements and fragmentation complicated gene calling, 15,419 gene models were predicted, laying the foundations for subsequent work. The authors in conclusion noted the limitations imposed on future research by the fragmentary nature of the genome and recognized filling gaps in the assembled sequence would need to be a high priority.

Given the poor mapping for the original assembly, and with the continued failure of polytene chromosome preparation, new assembly techniques were needed. Between 2008 and 2010, additional work on chromosomal physical mapping occurred, with Fan Yang and Maria Sharakhova identifying the imaginal discs of 4th instar larvae as an abundant source of chromosomes with lengths 4-5 times longer at prometaphase than at the more commonly used metaphase. When combined with cold temperature preparation in hypotonic solutions, these provided increased resolution for FISH mapping (Yang 2010, Sharakhova et al. 2011). Using these improved chromosomal preparations, an integrated linkage, chromosome and genome map was finally developed, involving the physical mapping of 100 bacterial artificial chromosome (BAC) clones, and the anchoring of 183 Mbp (13.3 % of the genome) to exact positions on *Ae. aegypti* chromosomes (Timoshevskiy et al. 2013). This was extended by an additional 368 BAC clones the following year, locating 624 Mbp or 45% of the genome to chromosomal bands, enabling syntenic comparisons with *An. gambiae* (Timoshevskiy et al. 2014). These mapping projects enabled steady improvements to the genome assembly, leading to the release of the *AaegL2* assembly (GenBank assembly GCA_000004015.2). In the subsequent year, the use of restriction-site associated DNA (RAD) sequencing led to an improved genetic map developed from the Liverpool substrain LVP-IB12 along with the COSTA RICO strain, built from over 2,000 markers (Juneja et al. 2014). This enabled 60% of the gene sequence to be assigned to chromosomes and scaffolds to be placed in approximate order. These mapping features were integrated into the *AaegL3* assembly in 2014 (GenBank assembly GCA_000004015.3).

### 4.1 Population genetics – Genome-wide era

Population genetics research in the post-genomic era can be divided into phases based on the dominant marker technology and the quality of supporting genomic resources. The first major consequence of the release of the draft genome was the increase in utility of microsatellite markers. Microsatellites were first developed in 2001 (Huber et al. 2001, Ravel et al. 2001), and have continued in use over the past 25 years. Original microsatellites had low variability, and few studies were published until 2007, when the newly published genome enabled more efficient bioinformatic searches for suitable markers (Nene et al. 2007). From 2007 to 2011, a larger number of microsatellite markers were developed (Chambers et al. 2007, Slotman et al. 2007, Endersby et al. 2009, Lovin et al. 2009, Brown et al. 2011). This led to a boom in microsatellite studies during the 2010s (Figure 1).

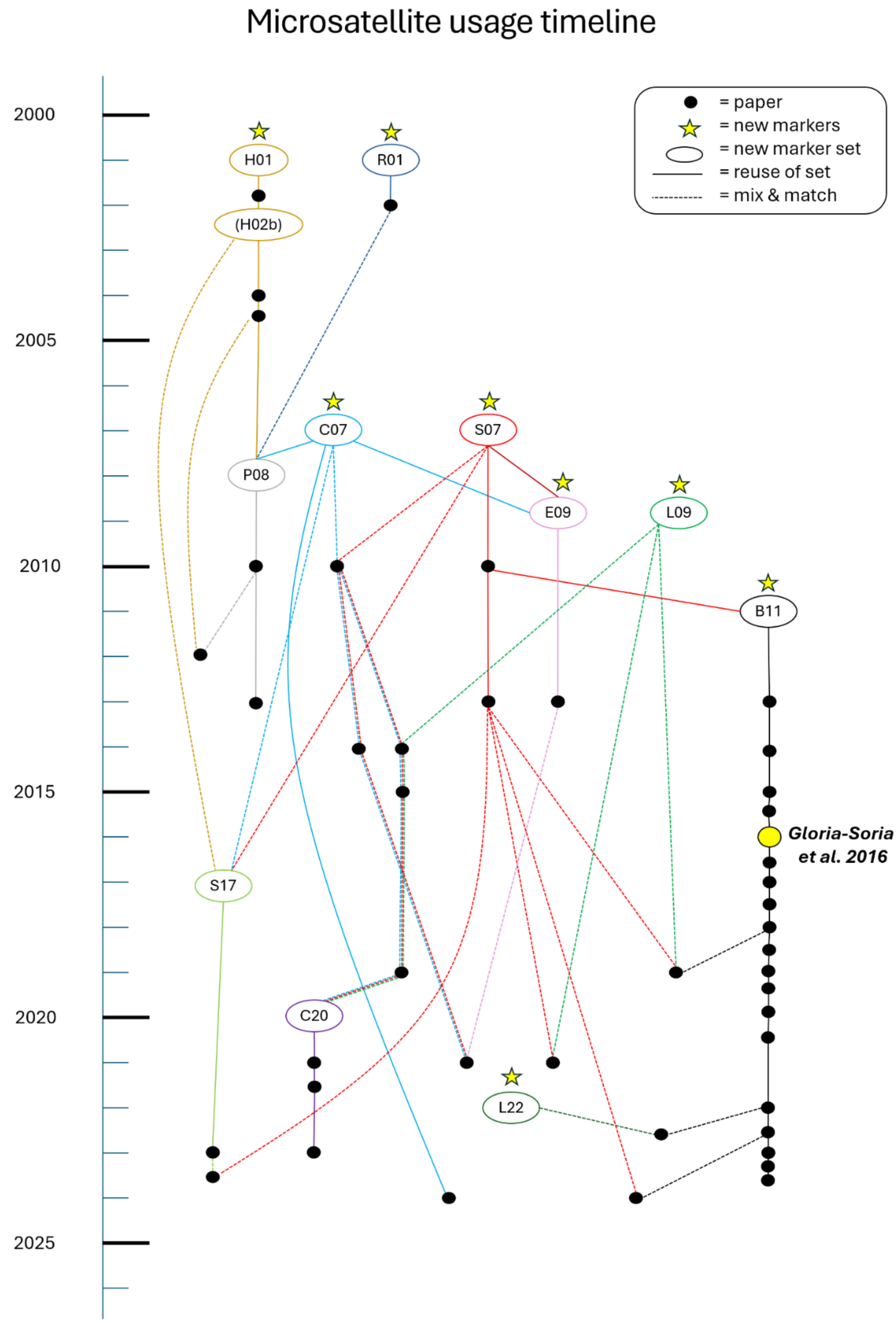

Figure 1. Microsatellite marker construction and re-use across 25 years of research in *Ae. aegypti*. Papers are shown as points. Papers used as sources for later studies are shown as ovals. Solid lines show whole or partial reuse from single source. Dashed lines show mix and match from multiple sources. Parallel dashed lines connect studies that mix and match from the same sources, but without choosing the same markers. Development of new microsatellite markers is indicated with a star. H01: (Huber et al. 2001); R01: (Ravel et al. 2001); H02b: (Huber, Luu Le Loan, et al. 2002); C07: (Chambers et al. 2007); S07: (Slotman et al. 2007); P08: (Paupy et al. 2008); E09: (Endersby et al. 2009); L09: (Lovin et al. 2009); B11: (Brown et al. 2011); S17: (Shi et al. 2017); C20: (Carvajal et al. 2020); L22: (Labiros et al. 2022). A key highlighted global microsatellite study is (Gloria-Soria et al. 2016). Full source list for figure: (Huber et al. 2001, 2004, Ravel et al. 2001, 2002, Huber, Luu Le Loan, et al. 2002, Paupy et al. 2004, 2008, 2010, 2012, Chambers et al. 2007, Slotman et al. 2007, Endersby et al. 2009, Lovin et al. 2009, Hemme et al. 2010, Hlaing et al. 2010, Brown et al. 2011, Brown, Obas, et al. 2013, Marcombe et al. 2013, Olanratmanee et al. 2013, Rasheed et al. 2013, Gloria-Soria et al. 2014, 2016, 2018, 2019, Mendonça et al. 2014, Monteiro et al. 2014, Rašić et al. 2014, 2015, Fonzi et al. 2015, Sayson et al. 2015, Calvez et al. 2016, Kotsakiozi et al. 2017, Pless et al. 2017, 2022, Shi et al. 2017, Kotsakiozi, Gloria-Soria, et al. 2018, Guagliardo et al. 2019, Maitra et al. 2019, Salgueiro et al. 2019, Seixas et al. 2019, Carvajal et al. 2020, 2021, Itokawa et al. 2020, Soghigian et al. 2020, Monsalve et al. 2021, Nirmani et al. 2021, Regilme et al. 2021, Edillo et al. 2022, Elnour et al. 2022, Labiros et al. 2022, Maynard et al. 2023, Muharromah et al. 2023, Rodrigues et al. 2023, Souza et al. 2023, Sumitha et al. 2023, Wang et al. 2023, Zhao et al. 2023, Abuelmaali et al. 2024, Mashlawi et al. 2024)

One problem with the use of microsatellites was that different studies added or removed loci from previous studies based on factors such as percentage of null alleles or the level of variability in the studied population. This reduced comparability in the literature as different combinations of microsatellites were used (Figure 1), though most later studies drew on microsatellites derived from the 2007 datasets (Chambers et al. 2007, Slotman et al. 2007). The exception to this general pattern was a 12 loci marker set developed in 2011 by Brown et al. (Brown et al. 2011) from a combination of eight of the C07 markers (Chambers et al. 2007 - Figure 1) with four additionally developed markers. This emerged as a de facto standard for the field, with the majority of studies from 2015 onwards using at least 11 of these markers. The compatibility of the markers within this subset of studies enabled broad integrative questions to be addressed, such as the global population structure of *Ae. aegypti* (Gloria-Soria et al. 2016), the identification of likely sources of recent range expansions (Gloria-Soria et al. 2014) or re-invasions (Kotsakiozi et al. 2017), the inferred origins of putative Mediterranean remnant populations (Kotsakiozi, Gloria-Soria, et al. 2018), and the likely sources of widely used laboratory strains (Gloria-Soria et al. 2019). The major phase of microsatellite study culminated with a landmark review of microsatellite-based population genetics and epidemiological records that used approximate Bayesian computation to argue for an initial West-African domestication of the species, followed by spread to the Americas and the Mediterranean driven by transatlantic shipping associated with the slave trade, then spread to the Indo-Pacific following the opening of the Suez Canal (Powell et al. 2018). One later study combined microsatellites and mitochondrial *CO1* haplotypes to examine the population structure of Southeast Asia and Australasia (Maynard et al. 2023).

The second phase of post-genomic population genetics came with the development of genome-wide SNP markers. The use of genome-wide SNPs began in 2013, when Brown et al. (2013) used RAD sequencing on 16 localities across Africa, the Americas, and the Asia-Pacific, identifying 1,504 biallelic SNPs following alignment to the *AaegL1* genome. An improvement came with the development of double digest RAD (ddRAD) sequencing using the cutting enzymes *NlaIII* and *MluCI,* which increased the number of SNPs accessible to the tens of thousands (Rašić et al. 2014). These datasets greatly increased the resolution relative to microsatellites in estimating population structure and fine-scale relatedness (Rašić et al. 2014). A follow-up study in the same system examined local population structure across the city of Yogyakarta, Indonesia, finding evidence that outlying regions are comparatively isolated from each other relative to the city centre. Additional studies using this ddRAD system established a lack of genetic evidence for the distinctness of *Ae. aegypti queenslandensis* from *Ae. a. aegypti* (Rašić et al. 2016), found extensive sex differentiated genetic regions along chromosome 1 (Fontaine et al. 2017), and used genomic measures of kinship to estimate intergenerational dispersal (Jasper et al. 2019), landscape dispersal barriers (Schmidt, Elfekih, et al. 2023), and loss of *Wolbachia* in full-sib families (Schmidt et al. 2018). Studies also investigated population structure in the Indo-Pacific (Schmidt et al. 2020) and the likely source of new *Ae. aegypti* invasive populations (Schmidt, Endersby-Harshman, et al. 2023). While ddRAD studies provide a large array of genome-wide variation for study, compatibility between datasets relies on a consistent choice of size selection window and digestion enzymes. A separate ddRAD project looked at population structure in the Caribbean (Sherpa et al. 2018); however, different restriction enzymes (*SbfI-HF* and *MspI*) were used, leaving this study incompatible with the studies of the previous group. A third series of studies of population genetics in Argentina again used differing restriction enzymes (*PstI* and *EcoRI*), leaving them incompatible with both previous sets of studies (Maffey et al. 2020, 2022).

ddRAD technology also provided the other major genome-wide SNP technology – the *Axiom_aegypti1* SNP chip, developed in 2015 and capable of genotyping 96 samples at a time across 50,000 SNP markers (Bennett et al. 2016). In the development process, the researchers genotyped 23 populations and 2 laboratory strains using a variety of methods including RNAseq and RAD, but the vast majority of SNPs (95%) were contributed by ddRAD. Unfortunately, these SNPs were also incompatible with prior studies: while the same restriction enzymes were used in this study as in Rašić et al. (2014), the size selection windows were different due to constraints in SNP chip design (195 ± 20 here vs 300-450 bp in the 2014 study). Of the 50,000 SNPs assessed, initial tests on field populations found 25,589 SNPs were usable for analysis. About 95% of SNPs in the dataset are ddRAD-derived. This SNP chip has been influential in ongoing population genetics studies, being further used to characterize the population structure and origin of *Ae. aegypti* into California (Pless et al. 2017, 2022), establish the origin of recently invasive populations (Gloria-Soria et al. 2018) and of a Black Sea remnant population (Kotsakiozi, Gloria-Soria, et al. 2018), to examine ancestral range population structure in Africa (Kotsakiozi, Evans, et al. 2018), to identify the sources of historic laboratory strains (Gloria-Soria et al. 2019), and to examine differences between mosquitoes in urban and forest habitats (Xia et al. 2020). The final genome-wide technology applied in this era was exon sequencing, which was used to examine West African *Ae. aegypti*, with demographic modelling via *fastsimcoal2* favouring Senegal *Aaa* populations being descended from the African precursors to the global *Aaa* invasion rather than being recently admixed from those populations (Crawford et al. 2017).

### 4.2 Resistance genetics – Genome-wide era

#### *4.2.1 VSSC (kdr)*

Following the initial identification of mutations at the *VSSC* gene subunit II, a subsequent study explored variation at this region using PCR and detected the mutations I1011V and V1016I (Saavedra-Rodriguez et al. 2007). The PCR primers used in this study were made available for future melting curve-based studies examining I1011/V1016. These primers quickly revealed V1016I in Brazil (Martins et al. 2009) and on Martinique (Marcombe et al. 2012). A retrospective study examining samples in Mexico from 1996 to 2009 found a steady, rapid increase in frequency from 0.04% before 2000 to 33.2% between 2006-2009 (García et al. 2009). The V1016I mutation was isolated from broader metabolic resistance via repeated back-crossing to the susceptible Rockefeller strain, and was shown to suffer fitness costs without selection (Brito et al. 2013). Following the detection of the F1534C mutation in the Caribbean (Harris et al. 2010), the distribution of both mutations was investigated in Brazil (Linss et al. 2014). The V1016I mutation was found to be a double mutant with F1534C, with F1534C also segregating independently. F1534C occurred across the country, while V1016I was less frequent in the north. A related study in Mexico re-examined historic samples from 2002-2012, finding a historical increase in V1016+F1534C frequency which was later eclipsed by an increase in frequency of the V1016I+F1534C double mutant (Vera-Maloof et al. 2015).

A further diagnostic assay was developed for V1016G, revealing this mutation was widespread throughout Thailand (Rajatileka et al. 2008). The V1016G mutation, in combination with S989P, were identified as candidates for deltamethrin resistance in the Khu Bua (Thailand) strain (Srisawat et al. 2010). A further mutation, D1794Y, was identified alongside this mutation in the Per-R resistant strain from Kaohsiung, Taiwan, following the sequencing of *VSSC* cDNA (Chang et al. 2009). In 2009, the first identification was made of the F1534C mutant in Thailand via sequencing of cDNA fragments (Yanola et al. 2008, 2010). This mutation was subsequently found at high frequency in Vietnam (Kawada et al. 2009), and then in the Caribbean, where it was strongly correlated to DDT and permethrin resistance (Harris et al. 2010). Later, F1534C was found alone and with V1016I in assays in Ghana (Kawada et al. 2016). Using intronic analysis, they suggested the V1016I mutant had spread from the Americas. A laboratory selection test found that V1016G out-competed F1534C against permethrin when both were present in the selected strain (Kasai et al. 2014), while *Xenopus* oocyte expression of *VSSC* found that S989P+V1016G sodium channels were less sensitive to permethrin and deltamethrin than F1534C, however the artificially produced S989P+V1016G+F1534C triple mutant was much less sensitive than any of the other mutation groups in isolation (Hirata et al. 2014). A later study showed that individuals heterozygous for these two mutations showed intermediate resistance to the two homozygous types (Plernsub et al. 2016).

### *4.2.2 GST Epsilon class*

Using recently developed expressed sequence tag databases, Nongkran Lumjuan conducted a fresh examination of the basis of metabolic resistance through GST-based DDT dehydrochlorination in resistant Thai strains as part of her PhD (Lumjuan 2005). She achieved this through searching the expressed sequence tag database using *Anopheles gambiae GST* sequences as queries, then cloning and sequencing identified products via RT-PCR, resulting in the initial identification of nine *GST* genes (Lumjuan et al. 2005). She additionally identified four amino acid substitutions found in the *GSTe2* gene of Thai resistant strains relative to the New Orleans strain: F115C, V150I, A178E and E198A. However, Lumjuan dismissed the significance of all but A198 on the basis of comparison to *An. gambiae*. Following the publication of the draft genomes, further expressed sequence tag matches were located in the whole genome shotgun database for *Ae. aegypti*, resulting in an identification of a total of 22 *GST* genes, including eight each from delta and epsilon classes (Lumjuan et al. 2007). Finally, DDT dehydrochlorinase activity was investigated for the three genes *GSTe2, GSTe5,* and *GSTe7* which were expressed at elevated levels in resistant Thai strains (Lumjuan et al. 2011). GSTe2 resistant and susceptible protein had similar activity, but the resistant protein had a higher affinity for DDT. GSTe2 and GSTe7 silencing also led to increased deltamethrin susceptibility, suggesting an additional role in pyrethroid resistance.

*4.2.3 Detoxification genes*

The draft genome enabled a genome-wide study of detoxification genes, which identified 235 detoxification genes, including 160 cytochrome P450s (CYP450s), 26 GSTs and 49 carboxy/choline esterases (Strode et al. 2008). The authors developed a microarray, the '*Aedes Detox Chip*' containing unique 70mer probes for 204 of the 235 detoxification genes. Tests on resistant strains from Thailand and the Caribbean found five significantly over-expressed genes shared between them: three *CYP9* P450s and the GSTe genes *GSTe2* and *GSTe7*. Most follow-up work using this chip focused primarily on the cytochrome P450 genes. A more detailed investigation of five field-sourced strains from Mexico and Peru following permethrin selection examined 290 genes on an expanded chip (Saavedra-Rodriguez et al. 2012). Very few genes were found to be consistently differentially expressed across lines relative to the New Orleans control. An important confounder in this study was the presence of the *V1016I* allele in the Mexican strains, which was found to reduce selection response in detoxification genes.

Notably, this paper did not consider the F1534C mutation. The authors mention 10 *CYP450* genes as upregulated, but pass over the *GSTe* and carboxy/choline esterase (CCE) genes found as well. Resistance was multifactorial, though the five generations of selection used was likely too few to develop a clear picture. A parallel study involving temephos selection also implicated *CYP450s* (*CYP6N12, CYP6J24)* and *GST*s (*AaGSTe2*) (Strode et al. 2012). Another *CYP450* gene, *CYP9J32*, featured in a comprehensive examination of pyrethroid resistance in the Caribbean, which examined both 1016I and 1534C pyrethroid resistance as well as with a transcriptomics microarray (Bariami et al. 2012). It was found to be upregulated in Grand Cayman and Cuba, and to have experienced a gene duplication. Following this, the *CYP450 CYP9J32* was expressed in *Escherichia coli* and shown to metabolize the pyrethroids permethrin and deltamethrin, as did three other *CYP450s* (Stevenson et al. 2012). The gene *CYP9J28* was transgenically expressed in *Drosophila melanogaster*, where it conferred pyrethroid resistance (Pavlidi et al. 2012).

Additional studies using RNAseq found both expression and polymorphism changes in P450 genes associated with selection (David et al. 2014, Kasai et al. 2014). Follow-up through introgression with the susceptible Rockefeller strain established that resistance-associated overexpression of five P450 genes was driven by *trans*-regulatory factors rather than by direct adaptation at their loci (Smith et al. 2018), and that the *VSSC* gene region contributed more to resistance than CYP-mediated pyrethroid resistance (Smith et al. 2019).

The first copy number variant (CNV) study was conducted using target enrichment in combination with high-throughput sequencing across populations from Thailand and South America resistant and susceptible to deltamethrin (Faucon et al. 2015). Out of 760 candidate genes, 41 resistance-associated gene amplifications were identified, mostly involving cytochrome P450s over-transcribed in resistant populations. This study also identified strong selection footprints in specific regions, highlighting 55 resistance-associated nonsynonymous SNPs. This study found *GSTe2* amplified approximately 2-fold. Note that at this time *GSTe2* was not yet mapped to a chromosome (this paper used the 2015 chromosomes maps of Timoshevsky and Juneja). The study was extended using RNA-seq, finding many CNVs had elevated expression (Faucon et al. 2017).

#### *4.2.4 Carboxylesterases* (*CCEs*)

A study of larval temephos resistance in the field from Thailand in combination with microarray analysis via the *Aedes Detox Chip* showed the CCE gene *CCEae3a* to be upregulated more than 60-fold in a strongly resistant population, and further investigation revealed a gene duplication at this site, as well as several amino acid polymorphisms (Poupardin et al. 2014). Later expression analysis in a baculovirus system found CCEae3a in *Ae. aegypti* was involved in temephos sequestration, but that amino acid polymorphisms did not affect kinetics (Grigoraki et al. 2016). The gene was later shown to produce resistance in a transgenic *An. gambiae* model (Poulton et al. 2024).

### 4.3 Challenges – Genome-wide era

Microsatellites and genome-wide SNPs led to a major improvement in our knowledge of the global population genetics of *Ae. aegypti*. However, the expanding number of papers published during this period also involved differ markers being used by different groups, with the datasets of one group incompatible with those of another. This limited options for reanalysing data of these different groups. Indeed, the transition from one technology to another led to equivalent challenges. The 11 microsatellite set developed by Brown et al. in 2011 (Brown et al. 2011) was within a few years regularly supplemented with and then superseded by the SNP chip developed by the same researchers (Evans et al. 2015). In more recent years, the data collected via the SNP chip has itself been challenged by the development of WGS sequencing and by the recognised potential for ascertainment bias (Gómez-Palacio et al. 2024).

**5 The whole-genome era: 2017-2026**

While slow improvements had occurred to mapping across the *L2* and *L3* genomes, in 2017, the application of the Hi-C approach led to the first assembly of chromosome-scale scaffolds in the *AaegL4* assembly (Dudchenko et al. 2017). This reused the original Sanger reads underlying all assemblies up until this point along with freshly generated Hi-C data to correct misjoins and merge scaffolds. While this genome project was in many ways primarily a stepping stone towards a full fresh assembly, it was immediately used to address biological questions that were sensitive to genomic position. Two separate studies in 2017 used this assembly to examine chromosome location and the *M* (male-determining) locus. The first used exon capture to study *Ae. a. formosus* in PK10, Senegal for sex-associated genes with increased heterozygosity in males. When compared to *Ae. a. aegypti*, PK10 *formosus* showed sex-associated genes along the length of the chromosome instead of clustered around the *M* locus, consistent with chromosomal rearrangements in this population (Campbell et al. 2017). The other study relied on ddRAD-based linkage mapping intercrosses within *Ae. a. aegypti* strains, and found a 123-Mbp region around the *M* locus with no detected recombination, and a smaller 63-Mbp region within it showing high male-to-female heterozygosity (Fontaine et al. 2017). This was interpreted as strongly challenging the presumption that this chromosome is pseudoautosomal.

The following year, a freshly sequenced, long-read and Hi-C-based assembly and annotation produced the *AaegL5* genome assembly (Matthews et al. 2018). In this assembly, the overall number of contigs was decreased by 93%, and contig N50 increased from 1.9 to 11 Mbp. 65% of the genome was composed of transposable elements. An amplification of *GSTe* genes was found in the assembly, as well as signatures of a selective sweep around the *VSSC* gene. The *AaegL5* reference genome was sequenced from the Liverpool strain; in 2022 a second genome assembly of another important historical *Ae. aegypti* strain, Rockefeller, was released (Fisher et al. 2022). This was deemed necessary as the Rockefeller strain is often used as a susceptible strain in resistance experiments.

Long-read and Hi-C technology contributed to genomic understanding in this era beyond just genome assembly. Linked read sequencing detected nine large inversions and 23 microinversions in representative global populations of both subspecies of *Ae. aegypti* as well as *Ae. mascarensis*, which were validated using breakpoint re-assembly or long-read alignment (Redmond et al. 2020). A major Hi-C proximity ligation study of 25 strains of *Ae. aegypti* as well as one of *Ae. mascarensis* revealed 21 multimegabase polymorphic inversions ranging from 5 to 55 Mbp (Liang et al. 2025). Fifteen inversions were Africa-specific, and three specific to non-African populations. Inversions were validated with FISH. This study is very close to a whole-genome assembly, as the (pooled) Hi-C lines were sequenced to 50X coverage across the genome and Hi-C information supplies genomic context specific to that strain. A study characterized transposable element structure in global *Ae. aegypti* genomes, finding they contribute strongly to genetic diversity and reflect similar population structure to that seen in SNPs (Valente-Almeida et al. 2025).

Finally, a marker-assisted mapping approach was developed, involving the incorporation of an EGFP (fluorescent) transformation marker into the *Ae. aegypti* genome (Chen et al. 2022). Using the technology, in 2022 researchers were able to identify the source of the historic *re* locus' phenotype of red eye to be the gene *cardinal*. This represents a full circle via modern genomic technology to one of the initial variants discovered by McClelland, which, as part of the *triple-marked strain,* supported almost all the genetic and physical mapping ever since, including the sequencing of the Liverpool genome.

### 5.1 Population genetics – Whole-genome era

While important work continued with genome-wide sequencing studies which now had a continuous assembly for alignment, the first whole genome resequencing (WGS) population genetic study occurred in 2019 with the examination of the population structure of Californian *Ae. aegypti* (Lee et al. 2019). In total, 46 individuals were sampled from California, four from Florida and three from South Africa. 4.7 million high-quality SNPs were detected – an increase of more than an order of magnitude from previous genome-wide studies. In contrast to the broader experience with genome-wide SNP assays and microsatellites, WGS studies, notionally covering the entire genome, were intrinsically compatible, enabling future studies to build on the datasets of prior studies. Indeed, in a follow-up study, they added in an additional 38 genomes from populations across the United States (Kelly et al. 2021). This dataset was later repurposed to identify loci potentially involved in local adaptation via integration with topological and climate variables (Soudi et al. 2023).

A substantial step forward was taken with the examination of 375 mosquitoes from sites across Africa in a dataset that included information about odour preference for human vs animal (Rose et al. 2020). Odor preference was correlated with dry season intensity and human population density. Human preference outside of Africa was identified as having a single genomic origin, with the *Aaa* ancestry component from an Admixture analysis with K=3 correlated with white scales and human preference. The study supported the hypothesis of a Sahel origination of the globally invasive *Aaa* subspecies. In a follow-up study using the same dataset, researchers applied the Multiple Sequentially Markovian Coalescent (MSMC) (Schiffels and Wang 2020) and a related program MSMC-IM to examine the divergence between generalist and human-specialist *Ae. aegypti* populations in West Africa. Calibrating the analysis with historical spread of *Ae. aegypti* to the Americas, they found the divergence between specialist and generalist populations occurred approximately 5,000 years ago. Other researchers added an additional 52 *Ae. aegypti* from global populations and explored the genomic adaptive basis of the *Aaf* to *Aaa* transition, identifying adaptive signals in 186 genes, many of which arose from standing variation in *Aaf* populations (Lozada-Chávez et al. 2025).

This series of WGS studies culminated in 2025, when 1,206 genomes were published as a single dataset. New individuals were predominantly sourced from the Americas, including the Caribbean, and used to explore the spread and diversification of global *Aaa* populations (Crawford et al. 2025). The study examined the species in three genetic clusters: *Aaf*, which corresponds to the African generalist lineage; proto-*Aaa*, which corresponds to African human specialist and any genetically similar populations, and *Aaa*, which corresponds to globally invasive *Aaa*. This study estimated split times between *Aaf* and proto-*Aaa* of 2,300 years ago, between African and South American proto-*Aaa* of approximately 320 years ago, and between proto-*Aaa* and invasive *Aaa* of approximately 220 years ago. A contrasting picture of the specific origins of North America *Ae. aegypti* was argued by Kent et al. (Kent et al. 2025), who used previously sequenced global samples to propose multiple introductions of *Ae. aegypti* to North America, based primarily on ADMIXTURE results and inferring historic population sizes and split times using SMC++. They argue that California has experienced multiple historic introductions that have remained genetically separate over hundreds of thousands of years, and that both lineages are found across the state in close proximity to each other. However, they do not account for the explicit model tests in Rose et al. (2020) and Crawford et al. (2025) which include the same samples.

### 5.2 Resistance genetics – Whole-genome era

#### *5.2.1 VSSC (kdr)*

The whole-genome era saw a continued focus on tracing the spread of known mutants of the *VSSC* gene through PCR assays. In addition, *Xenopus* expression experiments examined the comparative role of V1016I, V1016G, T1520I and F1534C in resistance to type I (permethrin) and type II (deltamethrin) pyrethroids (Chen et al. 2019). The authors suggest F1534C emerged first, with V1016I and T1520I emerging later as double mutants. Further investigations crossbreeding resistant with susceptible strains identified fitness costs associated with the S989P+V1016G double mutant (Rigby et al. 2020), as well as with the V*410L+F1016I+V1534C* allele (Silva et al. 2023).

Some *VSSC* studies used oligomer probes to enable targeted sequencing of the majority of *VSSC* exons, which are sparsely distributed across almost 500 kbp of the genome (Itokawa et al. 2019), or targeted amplicon approaches testing multiple resistance genes (Collins et al. 2022). Additionally, some genotyping studies began to take into account the intronic regions adjacent to mutation-containing exons, which enabled initial work on their broader haplotypes (Chung et al. 2019). Two such studies separately argued that the F1534C mutation had originated multiple times (Cosme et al. 2020, Fan et al. 2020). Both studies relied on intronic segments: Fan et al. (2020) on segments near each end of the 450 kbp gene, and Cosme et al. (2020) on introns adjacent to the V1016I and F1534C mutations (which are approximately 45 kbp apart). Two haplotype clades were found near the V1016 position. However, these studies did not consider the possibility of recombination across regions between 45 and 450 kbp, which could produce similar signals to if there had been multiple independent substitutions.

*VSSC* mutant information was additionally integrated with broader population genetics via ddRAD-based markers to identify the geographical origin of airport incursions of *Ae. aegypti* in Australia as well as their resistance alleles (Schmidt et al. 2019). A follow-up study achieved an important integration between resistance and demographic population genetics in showing that variation at SNPs within the *VSSC* gene was structured by resistance profile within the Indo-Pacific, in contrast to genome-wide SNPs which were structured by population (Endersby-Harshman et al. 2020).

### *5.2.2 Metabolic resistance*

Contemporaneous to the above work, the connection between *VSSC* substitutions and broader metabolic resistance to deltamethrin was examined using exon sequencing of *Ae. aegypti* from Mexico (Saavedra-Rodriguez et al. 2019). Authors noted a strong association of deltamethrin resistance with the *VSSC* region on chromosome 3, as well as various cytochrome P450s and esterases with weaker associations. However, a consequence of using the *AaegL5* genome for sequence alignment was that they inadvertently removed all information about key *GSTe* class genes, as these are subject to a duplication in the *AaegL5* genome which interferes with naïve mapping approaches (Matthews et al. 2018). The same researchers conducted a follow-up study involving isofemale lines, but again, after aligning to the *AaegL5* genome, lost the signal from the *GSTe* cluster (Saavedra-Rodriguez et al. 2021).

Another study used a hybrid exon-capture and genome-wide probe DNA-seq approach to identify 39 resistance-associated CNVs in pooled $F_2$ crosses, covering a variety of genes including *CYP6* and *CYP9* cytochrome P50s, carboxylase, and some *GST* genes (Cattel et al. 2020). A follow-up study identified several gene amplifications affecting the carboxylase cluster segregating in populations in Southeast Asia (Cattel et al. 2021). While genome-wide SNPs identified a selective signature associated with the *GSTe* cluster in these studies, mapping to the *AaegL5* genome again meant that no genes of this family were tested for presence of CNVs. In practice, this means that the *AaegL5* genome, while broadly an improvement on its predecessors, and while including a discussion of the *GSTe* cluster in its release paper (Matthews et al. 2018). may have led to a neglect of this form of resistance in the whole-genome era.

The gene *GSTe2* was, however, investigated via expression in a temephos resistant/susceptible strain pair originating in Brazil measured for GST activity and expression then cloned and expressed in *E.coli* (Helvecio et al. 2020). Researchers found increased expression in the resistant strain and identified variants within this gene which were confirmed to have a higher catalytic activity as measured by the CDNB substrate. However, this paper does not consider gene amplification, even though the resistant strain showed increased expression levels. Further, the authors made two crucial assumptions: first, that the *AaegL5* reference genome contains the wild-type sequence; and second, that an increase in catalytic effect against a generic measure of GST activity has a correlation to specific catalysis against insecticides such as temephos. Activity against insecticides was not tested. A further study investigated and did not find *GSTe2* association with resistance in a knockdown assay, but this was likely underpowered – only 21 individuals were typed for *GSTe2* mutations and background *VSSC* genotypes varied (Kampango et al. 2022).

A study combining RNA-seq and whole genome pool-seq identified a 220 kbp genomic duplication covering a cytochrome CYP6 P450 cluster in isofemale deltamethrin resistant lines from French Guiana which had been crossed to remove most major *VSSC* mutants (Bacot et al. 2024). This paper also found a partial duplication associated with the 3' end of the *VSSC* gene, with the duplicated region being non-functional and containing wild-type codons. Breakpoints were identified for both duplications using long-read sequencing. A Cas9 susceptible strain was used to repeatedly cross to a highly permethrin-resistant Puerto Rico strain to examine detoxification-based resistance with transcriptomic profiling, identifying more than 70 differentially expressed genes (Brown et al. 2026). Note that transcriptomic (RNA-seq) reads were “aligned to the *Ae. aegypti* reference genome (*AaegL5_3*).” This almost certainly means that core *GSTe* genes failed to align.

### 5.3 Population genetics of resistance – Whole-genome era

Finally, studies began to incorporate WGS datasets for exploration of insecticide resistance, beginning with an examination of sequences from Colombia (Love et al. 2023). Through combining these newly sequenced individuals with prior datasets from the Americas and Africa, they used the program SweepFinder2 to identify three high confidence selective sweeps in Americas populations: one between 351-352 Mbp on chromosome 2 (the position of the *GSTe* genes) shared by Brazilian and Colombian samples; one between 356.3 and 356.6 Mb in Brazil only, overlapping the cytochrome gene *CYP4H31*; and the third, shared by both, at the location of the *VSSC* gene on chromosome 3. This study also examined nucleotide diversity and Tajima's D at the *VSSC* locus. A further WGS analysis of global selection in 2024 used 729 global *Ae. aegypti* WGS samples available by that time and examined variation in the resistance genes *ace-1, GSTe2, rdl* and *Vssc* (Spadar et al. 2024). This enabled the construction of a catalogue of mutants at these genes within the 15 countries included in the study.

Another global study built on ddRAD-based genome-wide SNPs combined the above approaches, performing genome scans for selection-associated signatures as well as PCR-based genotyping at key *VSSC* sites in 16 countries across the Indo-Pacific as well as three global populations (Schmidt et al. 2024). This study found separate selective sweeps associated with Indo-Pacific and Americas F1534C mutations and the V1016G mutation, as well as multiple selective sweeps in the region of the *GSTe* gene cluster. However, the limitations of ddRAD data prevented further exploration of the genomic basis of these sweeps. This paper represents an incorporation of both demography and selection in examining the population genetics of *Ae. aegypti*.

### 5.4 Challenges – Whole-genome era

While the *AaegL5* genome was overall a vast improvement on other genomes, enabling context-aware treatment of genomic variants across multiple marker types, and especially powering WGS-based population genomics, the identification of duplicated *GSTe* genes is a major oversight that may have set back an understanding of the contribution of the *GST* gene cluster to selection. Almost every major study examining resistance alleles since 2017 that positions *GST* alleles on the genome, either for CNV detection or examination of selective sweeps, has at some point aligned its variants naively to the *AaegL5* genome and lost the core *GSTe* genes. As the number and breadth of those papers has grown, in stark contrast to prior eras, the *GST* genes have likely been increasingly marginalized in the literature, with their place taken by other metabolic resistance families and the target site genes. It is possible that other genes have also been duplicated in *AaegL5*; if so, their study will likely involve the same difficulties described here.

## 6 Towards a pangenome era

Two papers released in the past year potentially herald a new phase of genomics in *Ae. aegypti*. Up until 2025, only the Liverpool strain and then later the Rockefeller strain were sequenced (Nene et al. 2007, Matthews et al. 2018, Fisher et al. 2022). However, in 2025, researchers using PacBio HiFi sequencing with Hi-C sequencing produced *de novo* genome assemblies of *Ae. aegypti* from Burkina Faso and of its sister species *Aedes mascarensis* (Morinaga et al. 2025). This paper notes two issues possessed by the *AaegL5* genome that have since been overcome: (1) the requirement of high DNA input for library construction, and (2) the error-proneness of long read sequencing in the 2010s. The advent of highly accurate long reads directly reduced the requirement for large amounts of DNA from a single individual. While this paper benefited from the overcoming of substantial technical issues, it is less clear whether it sufficiently considered population genomic issues: the *Ae. aegypti* from Burkina Faso was sampled in Ouagadougou in October 2021. This West African city population has shown relatively higher preference for human odour than rural *Aaf* populations (Rose et al. 2020), which has been associated with a recent influx of human-specialist ancestry (Rose et al. 2023); it is not clear whether it is truly representative of the *Aaf* subspecies. Nonetheless, the genome itself is a valuable contribution to understanding this species. The next year a follow-up study sequenced two additional genomes using the same methods: wild-caught *Aaa* from North America, and wild-caught *Aaa* from Southeast Asia (Morinaga et al. 2026 Apr 22). These assemblies could in principle be repeated from a variety of other global populations and substantially improve our overall knowledge of the population genomics of the species.

## 7 Molecular genetics in *Ae. aegypti* – new directions and oversights

Several themes can be drawn from a consideration of the history of genomic mapping and markers in *Ae. aegypti* in relation to population genetics and the genetics of resistance.

### 7.1 Genome assemblies

*Ae. aegypti* is now in an extremely strong position for genomics research, with, as of 2026, four published whole genome assemblies as well as the recent publication of the genome of the sister species *Ae. mascarensis* (Morinaga et al. 2025). This opens up the possibility of increased population-wide characterization of structural variants, enabling a fuller picture of diversity and ongoing evolutionary history to be developed (Sharakhova and Sharakhov 2025), as well as facilitating pangenomic (Eizenga et al. 2020) and ancestral recombination graph-based analyses (Speidel et al. 2021, Grundler et al. 2025).

However, the features that held back early chromosomal work within the species are still present and exert impacts on typical WGS datasets today. As the majority of the genome consists of repetitive elements, it is difficult for short-read datasets to achieve well-phased haplotypes over lengthy regions. In addition, a genome with high transposable element content may be more prone to small-scale chromosomal rearrangements, meaning the immediate genomic context of gene in an individual population may not be the same as the same gene considered within a sequenced laboratory population where evolutionary changes in inversions may have taken place. While standard WGS studies struggle to deal with transposable elements, alternative approaches are being developed which use these regions to supplement characterizations of population genetic diversity and even seek out transposable elements with roles in adaptive evolution (Valente-Almeida et al. 2025).

**7.2 Population genetics – new directions and oversights**

Molecular population genetics in *Ae. aegypti* began with the consideration of sympatric types in Rabai, Kenya (Scott and Mcclelland 1975). Samples from the same approximate locations have now been incorporated into WGS datasets (Rose et al. 2020). However, in the interim between these collections, three separate genotyping technologies have come and gone, each eclipsing (and incommensurable with) the one before. Uncertainties within earlier datasets mean that it is only comparatively recently that the core features of the *Ae. aegypti* emergence and global spread have been widely accepted, and many details remain to be explored. While the global picture put forward by microsatellites (Powell et al. 2018) has many affinities with the current picture from WGS datasets (Crawford et al. 2025), the latter achieves a level of detail in consideration of split times and admixture that far exceeds anything possible before. Yet while WGS datasets have provided a mostly comprehensive coverage of Africa and the Americas, there are large unsampled regions across the Indo-Pacific that have not yet been integrated into these analyses. Crawford et al. in 2025 (Crawford et al. 2025) covered 15 sites from South America and 30 from Africa, but only one from the Pacific Islands and a handful from Southeast Asia. Increased coverage of this region is needed to add weight to population genetic assessments of *Ae. aegypti* invasion and subsequent population structure.

**7.3 Resistance genetics – new directions and oversights**

Resistance research began with the mapping of DDT resistance in larvae from strains on two continents to a locus on chromosome 2 and proceeded with the identification of a second locus on chromosome 3, as well as the dieldrin resistance locus. Today, genome sequencing, QTL mapping and transcriptomics have substantially increased our understanding of the complexity of resistance evolution in *Ae. aegypti*. The latter two resistant loci have been identified as the *VSSC* gene (cross-resistance to pyrethroids and DDT) and the *GABA* gene (organophosphate resistance). However, consideration of the original resistance locus has become part of the broader question of metabolic resistance, which as a default is studied from the perspective of polygenic interactions. No attempt has been made to identify the source, metabolic or otherwise, of what was initially understood as a single Mendelian factor with strong resistance effect. An assessment of the early literature with respect to current studies is also challenging as large gaps have been left in the transition from genetic map to genetic map as marker type changed. There is a need for more work to be done in establishing connections between historic genetic maps and current genome positions. The recent Cas9-confirmed marker-assisted mapping of the *re* mutant to *cardinal* is a promising move in this direction (Chen et al. 2022).

While the identify of $R^{DDT}$ on chromosome 2 has not been proven, several indications suggest that it involves the *GSTe* cluster. First, early characterization of larval DDT resistance in *Ae. aegypti* found increased glutathione-dependent DDT dehydrochlorination (Brown 1967). While the importance of the *VSSC* gene grew in analyses over the years (1975), $R^{DDT1}$ was still seen as a major locus of DDT resistance and to involve dehydrochlorination likely via glutathione, even after *VSSC* was identified as the key site of pyrethroid resistance in 1983 (Malcolm 1983b). Some initial work was done exploring field resistance with a *GST* gene basis in the early 90s (Grant and Matsumura 1989, Grant et al. 1991), although it was hindered by an incomplete knowledge of this gene family. However, with the dominance of DNA-based methods and the cloning of target site genes in the 1990s both *GST* genes and $R^{DDT}$ fell into relative neglect. Other than the post-PhD work of Lumjuan (Lumjuan et al. 2011) who put forward the importance of *GSTe2* for DDT and even pyrethroid resistance, and a brief consideration of its expression in a resistant strain from Brazil (Helvecio et al. 2020), the small (eight gene) *GSTe* gene family has only been studied as part of large transcriptomics datasets, where it is primarily competing based on expression level against ~50 Cytochrome P450 genes. In many of these studies, one or other *GST* gene has fallen out as being under selection, but the framing of these papers as to the polygenic nature of metabolic resistance and the prominence of P450 genes has led to its general neglect.

To make these matters worse, the incorporation of gene amplifications within the *GSTe* cluster in the widely used *AaegL5* genome assembly (Matthews et al. 2018) has led to an almost universal ascertainment issue in contemporary resistance and CNV studies with a genomic component, as standard mapping tools essentially filter the *GSTe* genes as repetitive elements (Saavedra-Rodriguez et al. 2019, 2021, Cattel et al. 2020, 2021, Brown et al. 2026). However, even with the key genes eliminated from the datasets, two recent studies involving genome-wide selection scans have identified sweeps associated with the 351-351 Mbp chromosome 2 region in *AaegL5* where the *GSTe* gene family sits (Love et al. 2023, Schmidt et al. 2024). There is an urgent need to reassess the evidence associated with the *GSTe* genes, especially *GSTe2*, which has been recently shown to play a role in cross-resistance to both pyrethroids and DDT in other mosquitoes (Kouamo et al. 2021, Diallo et al. 2022).

## Conclusions

This paper has reviewed the history of molecular genetic research in *Ae. aegypti*, focusing on available markers and genomic maps for population genetics and on the identification of genetic bases of resistance. While overall the development of new genotyping techniques culminating in current WGS datasets has enabled increasingly sophisticated explorations of *Ae. aegypti* genomics, generational change between marker types and genetic maps has also resulted in a lack of direct comparability between older and more recent studies, requiring extensive re-sampling of the species' range with each iteration. This change has also led in some cases to the loss of an appreciation of the significance of earlier research questions. As gaps in research, we specifically identify the under-sampling of the Indo-Pacific region in the WGS genotyping era, as well as opportunities afforded by the intrinsic compatibility of WGS datasets. In addition, we draw attention to the neglect of the *GSTe* gene cluster and especially *GSTe2*, which presents a possible candidate for the first identified resistance site in *Ae. aegypti*. The *AaegL5* genome, while helpful, must be used cautiously when investigating signals of selection and CNVs due to the issues described here, and in critical cases should be examined closely.

## Methods

A narrative literature review was conducted using Google Scholar to identify research relevant to *Aedes aegypti* molecular genetics. Searches were initially conducted using the supplementary terms “resistance”, “genetics”, “genomics”, “population genetics” and “cytology” along with conceptual synonyms. Searches were conducted across the available publication period, with results examined chronologically to capture changes in the focus and development of the literature over time. The initial focus was on (1) shifts in genetic technology, and (2) advances in studies of resistance, population genetics and chromosomal structure.

Search results were screened initially by title, with the large majority of returned titles examined for relevance. Abstracts were subsequently reviewed for publications that appeared relevant to the scope of the review, and particularly relevant papers were examined in greater detail. Where an area emerged as relevant to the review, targeted searches were conducted using additional terms identified during the initial search and reading process. Citation networks were also followed in both directions, using reference lists to identify earlier work and Google Scholars "cited by" feature to identify later research. This iterative process was used to identify literature associated with themes that were not adequately captured by the initial search terms.

The literature was synthesized thematically rather than according to a predefined classification framework. Themes were developed iteratively during the search and reading process as recurring research questions, approaches, findings and areas of disagreement became apparent. Publications were then considered in relation to these themes.